\documentstyle[aps,prb,twocolumn,epsf,floats]{revtex}
\draft
\begin{document}
\wideabs{

\title{The Fermion--Boson Transformation 
       in Fractional Quantum Hall Systems}

\author{
   John J. Quinn,$^{1,2}$
   Arkadiusz W\'ojs,$^{1,3}$
   Jennifer J. Quinn,$^4$ and 
   Arthur Benjamin$^5$}
\address{
   (1) University of Tennessee, Knoxville, Tennessee 37996\\
   (2) Oak Ridge National Laboratory, Oak Ridge, Tennessee 37831\\
   (3) Wroclaw University of Technology, Wroclaw 50-370, Poland\\
   (4) Occidental College, Los Angeles, California 90041\\
   (5) Harvey Mudd College, Claremont, California 91711}
\maketitle

\begin{abstract}
A Fermion to Boson transformation is accomplished by attaching to 
each Fermion a single flux quantum oriented opposite to the applied 
magnetic field.
When the mean field approximation is made in the Haldane spherical
geometry, the Fermion angular momentum $l_F$ is replaced by $l_B=
l_F-{1\over2}(N-1)$.
The set of allowed total angular momentum multiplets is identical 
in the two different pictures.
The Fermion and Boson energy spectra in the presence of many body 
interactions are identical if and only if the pseudopotential is 
``harmonic'' in form.
However, similar low energy bands of states with Laughlin correlations 
occur in the two spectra if the interaction has short range.
The transformation is used to clarify the relation between Boson and 
Fermion descriptions of the hierarchy of condensed fractional quantum 
Hall states.
\end{abstract}
}

\section{Introduction}
The transformation of electrons into composite Fermions (CF) by 
attaching to each particle a flux tube carrying an even number of 
flux quanta has led to a simple intuitive picture\cite{jain} of the
fractional quantum Hall effect (FQHE).\cite{prange}
Shortly after the introduction of the CF picture, Xie et al.\cite{xie}
introduced a Fermion$\rightarrow$Boson (F$\rightarrow$B) mapping 
connecting a 2D Fermion system at filling $\nu_F$ with a 2D Boson 
system at filling $\nu_B$, where $\nu_F^{-1}=\nu_B^{-1}+1$.
These authors stated that the sizes of the many body Hilbert spaces
for the Boson and Fermion systems were identical, and that their
numerical calculations verified that the mapping accurately transformed
the ground state of the Fermion system into the ground state of the
Boson system if and only if these ground states were incompressible FQH
states.
In this paper we show that the F$\rightarrow$B transformation leads to 
identical energy spectra if and only if the pseudopotential $V(L_{12})$ 
describing the interactions among the particles is of the ``harmonic'' 
form $V_H(L_{12})=A+B\,L_{12}(L_{12}+1)$, where $A$ and $B$ are constants, 
and $L_{12}$ is the total angular momentum of the interacting pair.
\cite{parentage1}
Laughlin correlations\cite{laughlin} occur when the actual pseudopotential
$V(L_{12})$ rises more quickly with increasing $L_{12}$ than $V_H(L_{12})$.
Anharmonic effects (due to $\Delta V(L_{12})=V(L_{12})-V_H(L_{12})$) cause
the interacting Fermion and interacting Boson spectra to differ for every
value of the filling factor $\nu_F=\nu_B(1+\nu_B)^{-1}$.
However, for appropriately chosen (short range) model pseudopotentials, 
the F$\rightarrow$B mapping accurately transforms the ground state of the 
Fermion system to that of the Boson system both for incompressible FQH 
states and for other low lying states.
The F$\rightarrow$B mapping is also very useful in understanding the 
relation between the Haldane\cite{haldane} hierarchy of Boson 
quasiparticle (QP) condensates and the CF hierarchy\cite{sitko,hierarchy} 
of Fermion QP condensed states.

\section{Gauge Transformations in Two-Dimensional Systems}
By attaching to each Fermion or Boson of charge $-e$ a fictitious flux
tube carrying an even number $2p$ of flux quanta oriented opposite to 
the applied field, the eigenstates and particle statistics are unchanged.
The ``gauge field'' interactions between the charge on one particle 
and the vector potential due to the flux quanta on every other particle 
make the Hamiltonian more complicated.
Only when the mean field approximation is made does the problem simplify.
In addition to these (CF and CB) transformations, a F$\rightarrow$B 
transformation can be made by attaching to each Fermion an odd number 
$2p+1$ of flux quanta (one flux quantum changes the statistics; other 
$2p$ flux quanta describe an additional CF or CB transformation).
If the particles are confined to the surface of a sphere containing 
at its center a magnetic monopole of strength $2S_F$ (for Fermions) 
or $2S_B$ (for Bosons) flux quanta, then the lowest shell of mean field
composite particles has angular momentum $l_F^*=|l_F-p(N-1)|$ where 
$l_F=S_F$ or $l_B^*=|l_B-p(N-1)|$ where $l_B=S_B$.
In the F$\rightarrow$B transformation (with $p=0$), $l_F$ is replaced
by $l_B=|l_F-{1\over2}(N-1)|$.

\section{Some Useful Theorems}
When a shell of angular momentum $l$ contains $N$ identical particles
(Fermions or Bosons), the resulting $N$ particle states can be 
classified by eigenvectors $\left|L,M,\alpha\right>$, where $L$ is the 
total angular momentum, $M$ its $z$-component, and $\alpha$ a label
which distinguishes independent multiplets with the same total angular
momentum $L$.
In the mean field CF (CB) transformation $l_F$ ($l_B$) is transformed
to $l_F^*$ ($l_B^*$).
In trying to understand why the mean field CF picture correctly predicted
the low lying band of states in the interacting electron spectrum, the
following theorem was important.\cite{parentage2}

\paragraph*{Theorem 1.}
The set of allowed total angular momentum multiplets of $N$ Fermions
each with angular momentum $l_F^*$ is a subset of the set of allowed
multiplets of $N$ Fermions each with angular momentum $l_F=l_F^*+(N-1)$.

Thus, if we define $g_{Nl}(L)$ as the number of independent multiplets
of total angular momentum $L$ formed by addition of the angular momenta 
of $N$ Fermions, each with angular momentum $l$, then $g_{Nl^*}(L)\le 
g_{Nl}(L)$ for every value of $L$.
A few examples for small systems suggest that the theorem is correct,
but a general mathematical proof is non-trivial.
A proof using the methods of combinatorics and the KOH\cite{ohara}
theorem has been given recently.\cite{quinn}
The same method allows the proof of a second theorem.

\paragraph*{Theorem 2.}
The set of allowed total angular momentum multiplets of $N$ Bosons each 
with angular momentum $l_B$ is identical to the set of multiplets for $N$ 
Fermions each with angular momentum $l_F=l_B+{1\over2}(N-1)$.\cite{quinn}

From Theorem 2 it follows immediately that Theorem 1 also applies 
to Bosons.
Theorem 2 is a stronger statement than a simple equality of the sizes
of the many body Hilbert spaces.\cite{xie}

\section{Interaction Effects}
In studying why the mean field CF picture correctly predicts the low 
lying states of a 2D electron system in a magnetic field
\cite{parentage1,parentage2} the ``harmonic pseudopotential'' 
\begin{equation}
   V_H(L_{12})=A+B\,\hat{L}_{12}^2
\label{eq1}
\end{equation}
was introduced.
Here $A$ and $B$ are constants and $\hat{L}_{12}$ is the total angular
momentum operator of the pair of particles.
It was shown that for the harmonic pseudopotential the energy of any 
multiplet of angular momentum $L$ was given by
\begin{eqnarray}
   E_{L\alpha}&=&A\cdot{1\over2}N(N-1)
\nonumber\\
              &+&B\cdot N(N-2)\,l(l+1)+B\cdot L(L+1).
\label{eq2}
\end{eqnarray}
The energy is independent of $\alpha$, so that every multiplet with
the same value of $L$ has the same energy.
Equation (\ref{eq2}) holds both for Fermions and for Bosons.
If $B_F=B_B=B$, then the spectrum of $N$ Bosons each with angular 
momentum $l_B$ is identical (up to a constant) to that of $N$ Fermions 
each with angular momentum $l_F=l_B+{1\over2}(N-1)$.
This is a rather surprising result because Fermions and Bosons sample
different sets of values of the pair angular momentum.
For example, for $N=9$ and $l_F=12$ (corresponding to $\nu_F={1\over3}$) 
the allowed values of the Fermion pair angular momentum consist of all 
odd integers between 1 and 23; for the corresponding Boson system with 
$l_B=8$ ($\nu_B={1\over2}$), the allowed values of $L_{12}$ are all 
even integers between 0 and 16.
Despite the totally different set of pseudopotential coefficients
sampled, up to a constant, the spectra of the Boson and Fermion 
systems interacting through a harmonic pseudopotential are the same.

In earlier work\cite{parentage2} it was emphasized that the harmonic
pseudopotential led to an ``anti-Hund's rule'' with the lowest energy 
state having the lowest allowed value of $L$.
It is the positive anharmonicity $\Delta V(L_{12})>0$ that causes
Laughlin correlations.
It is useful to introduce the ``relative'' angular momentum 
${\cal R}=2l-L_{12}$.
For Bosons ${\cal R}_B=0$, 2, 4, \dots\ while for Fermions 
${\cal R}_F=1$, 3, 5, \dots; in both cases, ${\cal R}\le2l$.
We can write the pseudopotential and its harmonic and anharmonic parts 
in terms of ${\cal R}$, and call them $V({\cal R})$, $V_H({\cal R})$, 
and $\Delta V({\cal R})$, respectively.
It is more reasonable to make simple models for $\Delta V({\cal R})$
(e.g. assume that it vanishes for all ${\cal R}$ greater than some 
value) than for $V({\cal R})$ itself.
From equation (\ref{eq2}) and the equation for the total energy,
\begin{equation}
   E_{L\alpha}={1\over2}N(N-1)\sum_{\cal R}
               {\cal G}_{L\alpha}({\cal R})\,V({\cal R}),
\label{eq3}
\end{equation}
where ${\cal G}_{L\alpha}({\cal R})$ is the coefficient of fractional
grandparentage (CFGP),\cite{parentage1,parentage2,parentage3} it is 
readily ascertained that the interacting Boson and interacting Fermion 
systems cannot have identical spectra when $\Delta V({\cal R})$ is 
non-zero.

Xie et al.\cite{xie} determined the Boson and Fermion eigenfunctions
by exact numerical diagonalization for six particle systems connected 
through the F$\rightarrow$B transformation.
They then transformed the Boson eigenfunctions into Fermion 
eigenfunctions by multiplying them by $\prod_{i<j}(z_i-z_j)$, 
as required by the B$\rightarrow$F transformation.
The overlap of these transformed Boson eigenfunctions with the exact
Fermion eigenfunctions was then evaluated.
The overlap was quite close to unity for incompressible quantum fluid
states when the full Coulomb interaction was used.
A similar result was obtained for a model short range interaction 
$H_1$ for which $V({\cal R})$ vanished for ${\cal R}>1$ and was equal 
to the Coulomb values at ${\cal R}=0$ (for Bosons) or at ${\cal R}=1$ 
(for Fermions).
However, when the interaction was approximated by $H_3$ for which 
$V({\cal R})$ vanished for ${\cal R}>3$ and was equal to the Coulomb 
values at ${\cal R}=0$ and 2 (for Bosons) or at ${\cal R}=1$ and 3 
(for Fermions), the overlap was considerably smaller.
The reason appears to be that for Fermions $H_3$ is subharmonic at 
${\cal R}=3$, while for Bosons it is (marginally) superharmonic in the 
entire range of ${\cal R}$, and that for a subharmonic pseudopotential 
Laughlin correlations are not expected to occur.\cite{parentage3}
By a subharmonic (superharmonic) behavior of $V({\cal R})$ at a certain 
value ${\cal R}_0$ we mean that $V({\cal R}_0)$ is larger (smaller) 
than a value $V_H({\cal R}_0)$ for which $V({\cal R})$ would be 
harmonic (i.e., linear in $L_{12}(L_{12}+1)$) in the range 
${\cal R}_0-2\le{\cal R}\le{\cal R}_0+2$.
We will later use an anharmonicity parameter $x$ defined as 
$x({\cal R}_0)=V({\cal R}_0)/V_H({\cal R}_0)$; for the $H_3$ 
interaction, $x(3)=1.3$ (for Fermions) and $x(2)=0.8$ (for Bosons).

We have evaluated numerically the eigenstates of an eight electron 
system at $2S_F=19$ to 23 (these states correspond to Laughlin 
$\nu_F={1\over3}$ states with zero, one, or two QP's) for a number 
of different pseudopotentials.
We have used the full Coulomb pseudopotential, $H_1$, $H_3$, $H_5$, 
and a model pseudopotential $V_x$ in which $V_x(1)=1$, $V_x({\cal R}
\ge5)=0$, and $V_x(3)=x\cdot V_H(3)$ is an arbitrary fraction $x$ of 
the ``harmonic'' value.
We perform the same calculations for eight Boson systems at $2S_B=12$
to 16 (here, $V_x(0)=1$, $V_x({\cal R}\ge4)=0$, and $V_x(2)=x\cdot 
V_H(2)$).

\begin{figure}[t]
\epsfxsize=3.40in
\epsffile{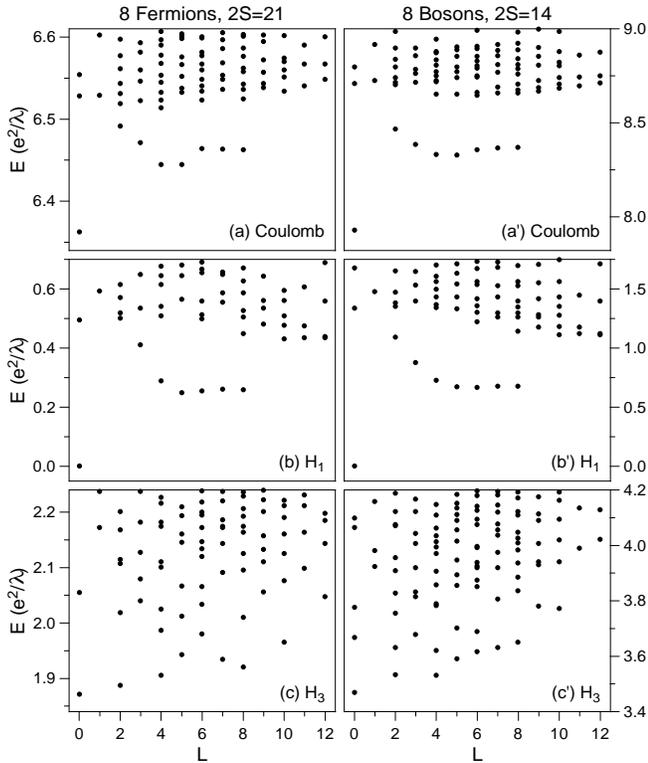}
\caption{
   The energy spectra (energy $E$ vs.\ angular momentum $L$) of 
   the corresponding eight Fermion (left) and eight Boson (right) 
   systems at the monopole strengths $2S_F=21$ and $2S_B=14$ 
   (filling factors $\nu_F={1\over3}$ and $\nu_B={1\over2}$) 
   for the Coulomb pseudopotential in the lowest Landau level 
   (a--a$'$), and for the model pseudopotentials $H_1$ (b--b$'$), 
   and $H_3$ (c--c$'$). $\lambda$ is the magnetic length.
}
\label{fig1}
\end{figure}

\begin{figure}[t]
\epsfxsize=3.40in
\epsffile{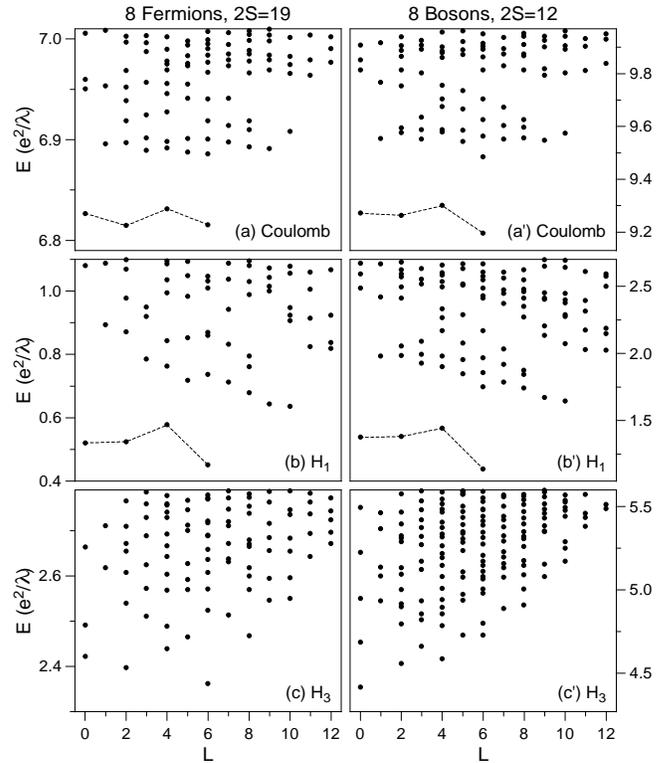}
\caption{
   The energy spectra (energy $E$ vs.\ angular momentum $L$) of 
   the corresponding eight Fermion (left) and eight Boson (right) 
   systems at the monopole strengths $2S_F=19$ and $2S_B=12$ 
   (two Laughlin quasielectrons in the $\nu_F={1\over3}$ and $\nu_B
   ={1\over2}$ state) for the Coulomb pseudopotential in the lowest 
   Landau level (a--a$'$), and the model pseudopotentials $H_1$ 
   (b--b$'$), and $H_3$ (c--c$'$). $\lambda$ is the magnetic length.
}
\label{fig2}
\end{figure}

In Fig.~\ref{fig1} we contrast the energy spectra for the Fermion 
and Boson systems at $\nu_F={1\over3}$ ($\nu_B={1\over2}$) for the
Coulomb pseudopotential appropriate for the lowest Landau level 
(a--a$'$), and for the model pseudopotentials $H_1$ (b--b$'$) and 
$H_3$ (c--c$'$).
In Fig.~\ref{fig2} we do the same for the state containing two 
Laughlin quasielectrons (QE).
The lowest states in (a--a$'$) and (b--b$'$) are quite similar 
consisting of a Laughlin $L=0$ ground state in Fig.~\ref{fig1} 
and two-QE states with $l_{\rm QE}={1\over2}(N-1)={7\over2}$ giving 
$L=N-2$, $N-4$, $\dots=0$, 2, 4, and 6 in Fig.~\ref{fig2}.
The magnetoroton band (at $2\le L\le8$) is apparent in Fig.~\ref{fig1}
although the gaps and band widths are different for different 
pseudopotentials.
The pseudopotential used in (c--c$'$) gives very different results 
both in Fig.~\ref{fig1} and Fig.~\ref{fig2}.
As mentioned before, this results because $V(3)$ used in Fermion 
pseudopotential $H_3$ is too large to lead to Laughlin correlations.

\begin{figure}[t]
\epsfxsize=3.40in
\epsffile{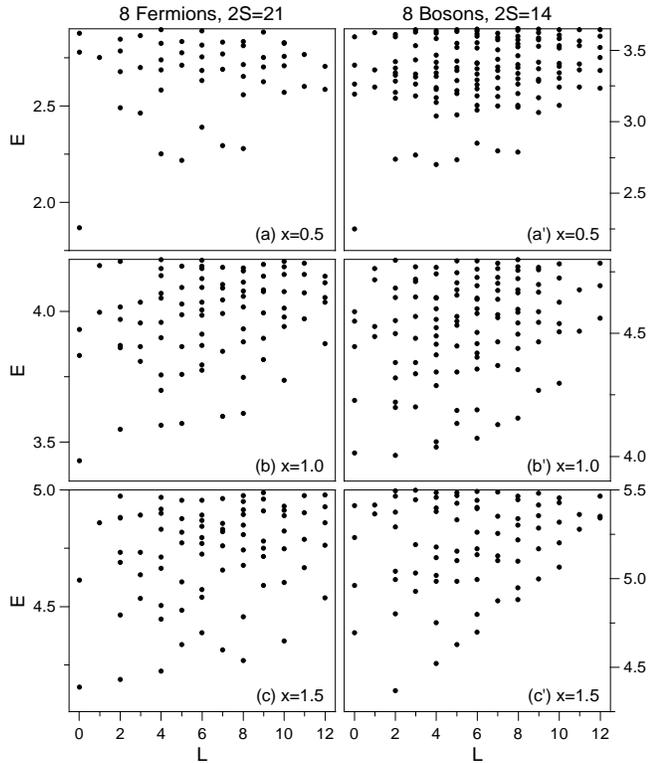}
\caption{
   The energy spectra (energy $E$ vs.\ angular momentum $L$) of 
   the corresponding eight Fermion (left) and eight Boson (right) 
   systems at the monopole strengths $2S_F=21$ and $2S_B=14$ 
   (filling factors $\nu_F={1\over3}$ and $\nu_B={1\over2}$) 
   for model interaction pseudopotentials $V_x({\cal R})$ with 
   $x={1\over2}$ (a--a$'$), $x=1$ (b--b$'$), and $x={3\over2}$ 
   (c--c$'$).
}
\label{fig3}
\end{figure}

To illustrate this point we have calculated the energy spectra using
pseudopotential $V_x$ with different values of $x$.
In Fig.~\ref{fig3} we show the spectra at $\nu_F={1\over3}$ ($\nu_B=
{1\over2}$) for $x={1\over2}$, 1, and ${3\over2}$.
For $x<1$, $V_x({\cal R})$ is superharmonic at ${\cal R}=3$ (for 
Fermions; $x\equiv x(3)$) or at ${\cal R}=2$ (for Bosons; $x\equiv 
x(2)$), and Laughlin correlations with an $L=0$ ground state occur.
For $x\ge1$ there is little resemblance between the numerical 
spectra and that associated with the full Coulomb interaction.
Furthermore, the Fermion and Boson spectra are quite different
from one another.

\begin{figure}[t]
\epsfxsize=3.40in
\epsffile{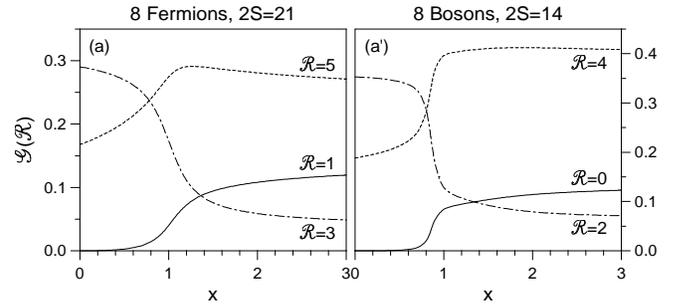}
\caption{
   The coefficients of fractional grandparentage ${\cal G}({\cal R})$
   from the pair states at three smallest values of ${\cal R}$ calculated 
   for the lowest energy $L=0$ state of the corresponding eight Fermion 
   (a) and eight Boson (a$'$) systems at the monopole strengths $2S_F=21$ 
   ($\nu_F={1\over3}$) and $2S_B=14$ ($\nu_B={1\over2}$) for the model 
   interaction $V_x$, as a function of $x$.
}
\label{fig4}
\end{figure}

From the eigenfunctions we can determine CFGP's ${\cal G}_{L\alpha}
({\cal R})$ for each state $\left|L,\alpha\right>$.
In Fig.~\ref{fig4} we plot the $x$-dependence of the CFGP's 
${\cal G}_{L\alpha}({\cal R})$ from pair states at three smallest 
values of ${\cal R}$ calculated for the lowest energy $L=0$ state 
of eight Fermions at $2S_F=21$ ($\nu_F={1\over3}$) and eight Bosons 
at $2S_B=14$ ($\nu_B={1\over2}$).
In both systems, a Laughlin incompressible state with vanishing 
${\cal G}(1)$ (for Fermions) or ${\cal G}(0)$ (for Bosons) occurs 
at small $x$, and a rather abrupt transition occurs at $x\approx1$,
implying a change of the nature of the correlations when the 
pseudopotential $V_x({\cal R})$ changes from super- to subharmonic.
At $x>1$, the correlations in the two systems are quite different and, 
for example, another abrupt transition occurs in the Boson system at 
$x\approx4$ (not shown in the figure), which is absent in the Fermion 
system.

\section{Quasiparticles}
The F$\rightarrow$B transformation allows us to better understand 
the Boson\cite{haldane} vs.\ Fermion\cite{sitko,hierarchy} description 
of QP's in incompressible FQH states.
Laughlin condensed states having $\nu_F=(2p+1)^{-1}$ (where $p$ is a 
positive integer) occur at $2S_F=(2p+1)(N-1)$ in the Haldane spherical
geometry.
The CF transformation\cite{jain} gives an effective angular momentum
$l_F^*=S_F^*=S-p(N-1)={1\over2}(N-1)$ when $2p$ flux quanta are attached 
to each electron and oriented opposite to the applied magnetic field.
Thus the $N$ CF's fill the $2l^*+1$ states of the lowest CF shell giving 
an $L=0$ incompressible ground state.

The F$\rightarrow$B transformation gives $2S_B=2S_F-(N-1)=2p(N-1)$
and a Boson filling factor of $\nu_B=(2p)^{-1}$.
Making a CB transformation gives $l_B^*=S_B^*=S_B-p(N-1)=0$.
This also gives an $L=0$ incompressible ground state because each
CB has $l_B^*=0$.
Thus the CF description of a Laughlin state has one filled CF shell
of angular momentum $l_F^*={1\over2}(N-1)$, while the CB description
has $N$ CB's each with angular momentum $l_B^*=0$.

For $2S_B=2n(N-1)\pm n_{\rm QP}$, where the $+$ and $-$ occur for
quasiholes (QH) and quasielectrons (QE), respectively, we define 
$2l_B^*=|2S_B^*|=n_{\rm QP}$.
This gives exactly the same set of angular momentum multiplets as
obtained in the CF picture with $2S_F=(2n+1)(N-1)+n_{\rm QH}$.
However it gives a larger set of multiplets than are allowed by 
$2S_F=(2n+1)(N-1)-n_{\rm QE}$.
For example, for $n_{\rm QE}=2$, $l_B^*=1$ and the allowed values 
of the pair angular momentum of the two QP's are $N$, $N-2$, $N-4$, 
\dots.
For a Fermion system with $l_F^*={1\over2}(N-1)-2$, the allowed
values of the QP pair angular momentum are $N-2$, $N-4$, \dots. 
The two sets can be made identical only if a hard core repulsion
forbids the Boson QP pair from having the largest allowed pair 
angular momentum $L_{12}^{\rm MAX}=N$.\cite{he}
This behavior is observed in Fig.~\ref{fig2} where the Boson 
treatment of two QE's (i.e., the CB transformation) would predict
states at $L=0$, 2, 4, 6, and 8, but the $L=8$ state does not 
occur in the low energy band.

Since the description of CB's (with hard core QE interaction) and 
CF's give identical sets of QP states, filled QP levels (implying 
daughter states) occur at identical values of the applied magnetic 
field.
In earlier work\cite{sitko,hierarchy} we have emphasized that both 
the Haldane hierarchy and CF hierarchy schemes assume the validity 
of the mean field approximation, and we have shown that this 
approximation is expected to fail when the QP--QP interaction is
subharmonic.
Numerical results show when the mean field approximation is
valid and when it fails.

\section{Summary}
We have shown that the F$\rightarrow$B transformation replaces the 
single Fermion angular momentum $l_F$ by $l_B=l_F-{1\over2}(N-1)$,
and that this transformation leads to an identical set of total
angular momentum multiplets.
The Fermion and Boson systems have identical spectra in the presence
of many body interactions only when the pseudopotential is harmonic,
i.e. linear in squared pair angular momentum, $L_{12}(L_{12}+1)$.
However, similar low energy bands of states with Laughlin correlations 
occur in the two spectra if the interaction pseudopotential is 
superharmonic, i.e. has short range.
We have studied numerically eight particle systems for different 
model interactions and shown the relation between the spectra and 
coefficients of fractional grandparentage for the Fermion and Boson 
systems.
Finally, we have used the F$\rightarrow$B transformation to clarify
the relation between the Haldane Boson picture and the CF picture
of the hierarchy of condensed states.

\section{Acknowledgment}
Research supported by the US Department of Energy under contract 
DE-AC05-000R22725 with Oak Ridge National Laboratory managed by
UT--Batelle, LLC.
John J. Quinn acknowledges partial support from the Materials Research 
Program of Basic Energy Sciences, US Department of Energy.

\end{document}